\documentclass[aps,twocolumn,showpacs,preprintnumbers,amsmath,amssymb,superscriptaddress,prr,longbibliography]{revtex4-1}
\usepackage{amsmath}
\usepackage{amsfonts}
\usepackage{mathrsfs}
\usepackage{graphicx}
\usepackage{times}
\usepackage{color}
\usepackage[normalem]{ulem}
\usepackage{epstopdf}

\newcommand\redout{\bgroup\markoverwith{\textcolor{red}{\rule[.5ex]{2pt}{1pt}}}\ULon}

\def\be{\begin{equation}}
\def\ee{\end{equation}}
\def\bea{\begin{eqnarray}}
\def\eea{\end{eqnarray}}

\begin{document}

\title{Localization-delocalization Transition in an electromagnetically induced photonic lattice}

\author{Rui Tian}
\affiliation{Ministry of Education Key Laboratory for Nonequilibrium Synthesis and Modulation of Condensed Matter,Shaanxi Province Key Laboratory of Quantum Information and Quantum Optoelectronic Devices, School of Physics, Xi'an Jiaotong University, Xi'an 710049, China}

\author{Shuai Li}
\affiliation{Ministry of Education Key Laboratory for Nonequilibrium Synthesis and Modulation of Condensed Matter,Shaanxi Province Key Laboratory of Quantum Information and Quantum Optoelectronic Devices, School of Physics, Xi'an Jiaotong University, Xi'an 710049, China}

\author{Maksims Arzamasovs}
\affiliation{Ministry of Education Key Laboratory for Nonequilibrium Synthesis and Modulation of Condensed Matter,Shaanxi Province Key Laboratory of Quantum Information and Quantum Optoelectronic Devices, School of Physics, Xi'an Jiaotong University, Xi'an 710049, China}

\author{Hong Gao}
\affiliation{Ministry of Education Key Laboratory for Nonequilibrium Synthesis and Modulation of Condensed Matter,Shaanxi Province Key Laboratory of Quantum Information and Quantum Optoelectronic Devices, School of Physics, Xi'an Jiaotong University, Xi'an 710049, China}

\author{Yong-Chang Zhang}
\affiliation{Ministry of Education Key Laboratory for Nonequilibrium Synthesis and Modulation of Condensed Matter,Shaanxi Province Key Laboratory of Quantum Information and Quantum Optoelectronic Devices, School of Physics, Xi'an Jiaotong University, Xi'an 710049, China}

\author{Bo Liu}
\email{liubophy@gmail.com}
\affiliation{Ministry of Education Key Laboratory for Nonequilibrium Synthesis and Modulation of Condensed Matter,Shaanxi Province Key Laboratory of Quantum Information and Quantum Optoelectronic Devices, School of Physics, Xi'an Jiaotong University, Xi'an 710049, China}

\begin{abstract}
We investigate the localization-delocalization transition (LDT) in an electromagnetically induced photonic lattice. A four-level tripod-type scheme in atomic ensembles is proposed to generate an effective photonic moir\'{e} lattice through the electromagnetically induced transparency (EIT) mechanism. By taking advantage of the tunable atomic coherence, we show that both periodic (commensurable) and aperiodic (incommensurable) structure can be created in such a  photonic moir\'{e} lattice
via adjusting the twist angle between two superimposed periodic patterns with square primitive. Furthermore, we also find that by tuning the amplitudes of these two superimposed periodic patterns, the localization-delocalization transition occurs for the light propagating in the aperiodic moir\'{e} lattice. Such localization is shown to link the fact that the flat bands of moir\'{e} lattice support quasi-nondiffracting localized modes and thus induce the localization of the initially localized beam. It would provide a promising approach to control the light propagation via the electromagnetically induced photonic lattice.
\end{abstract}

\maketitle

\section{Introduction}
Localized light can be used as a versatile tool for various manipulation and processing in optical information. It thus can be considered as one of the foundation for information dissemination. Past studies show that lots of promising methods, such as designing the optical localization propagation in optical fibers, utilizing artificial periodic structures in the photonic crystal and constructing random structures with Anderson localization effect, can implement the light localization \cite{akahane2003high, park2004electrically, joannopoulos2008molding, smith2004metamaterials, schurig2006metamaterial, han2014full}. In particular, one of the key ingredients of these schemes is to engineer the spatial characteristics of the optical medium, which shows unprecedented capabilities in controlling the flow of light as well as matter waves\cite{hu2005ultrafast,  zhang2008superlenses, bhandari1997polarization, lu2014topological, ozawa2019topological}.

Recently, another distinct approach to generate spatially periodic structures via the electromagnetically induced transparency (EIT) scheme \cite{ling1998electromagnetically}, either in hot atomic vapours \cite{sheng2015observation, zhang2018controllable, yuan2019integer} or ultracold atoms \cite{radwell2015spatially, yang2020dynamically}, has attracted considerable attention.
Many intriguing phenomena, such as  optical lattice solitons \cite{fleischer2003observation, michinel2006turning, zhang2011four}, photon-atom bound state \cite{longo2010few}, photonic Floquet topological insulators \cite{rechtsman2013photonic} and optical analogs of quantum random walks \cite{peruzzo2010quantum}, have been explored.

In this work, we propose a four-level tripod-type scheme in atomic ensembles to generate an electromagnetically induced photonic moir\'{e} lattice \cite{wang2020localization, fu2020optical} through the EIT mechanism. By taking advantage of the tunable atomic coherence, it is shown that the moir\'{e} pattern is highly flexible via changing the twist angle between two superimposed periodic patterns with square primitive. Both periodic (commensurable) and aperiodic (incommensurable) structure can be achieved.
Interestingly, we find a LDT of the light propagating in the aperiodic photonic moir\'{e} lattice, which manifests the typical flat-band feature of the moire lattice.

\section{Effective model}
\begin{figure}[b]
	\includegraphics[width=0.5\textwidth]{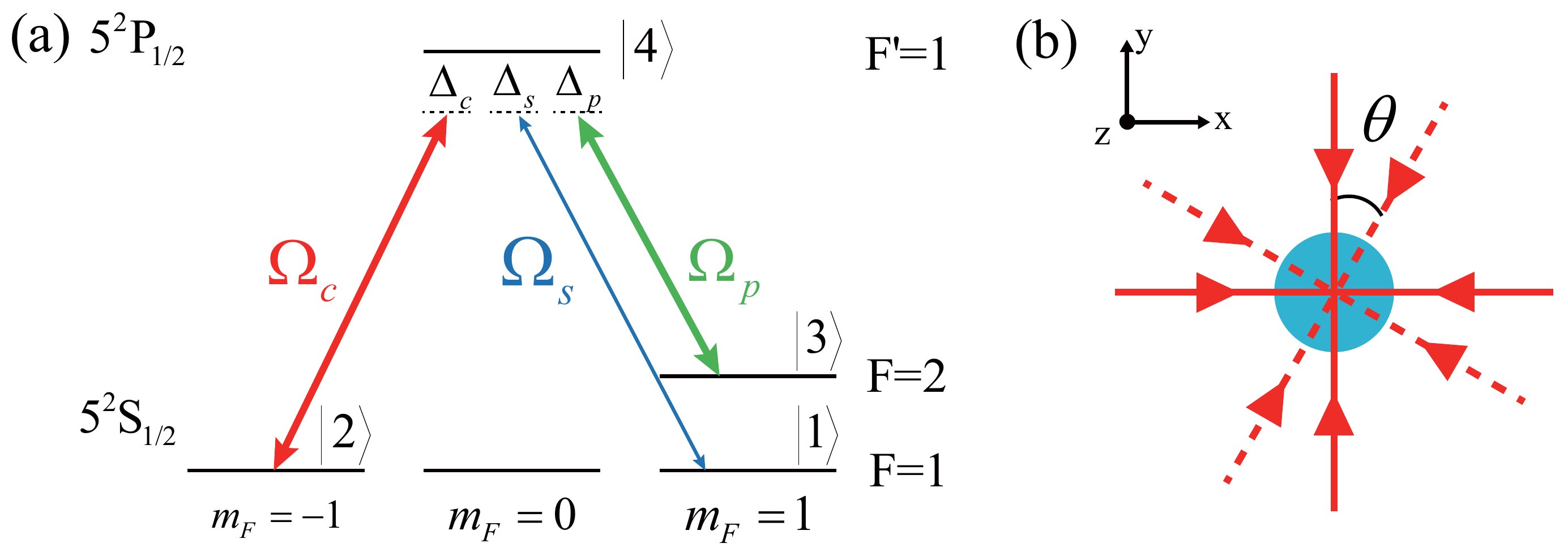}
	\caption{
		(a) The schematic plot of our proposed 4-level tripod scheme. Here $\Omega_{s(p,c)}$ stand for the the Rabi frequencies of signal, pump and coupling fields, respectively. $\Delta_{s(p,c)}$ labels the corresponding frequency detuning.
		(b)  Two groups of orthogonalized paired-standing waves marked by the solid and dashed lines, respectively, which can form two superimposed square patterns. $\theta$ labels the twisted angle between them.
	}
	\label{fig1}
\end{figure}

Let us take $^{87}\rm{Rb}$ atomic system as an example to show our proposed  four-level tripod-type scheme, which is
schematically presented in Fig.\ref{fig1}(a). The signal, coupling, pump fields drive the  transitions $|1\rangle\to|4\rangle, |2\rangle\to|4\rangle, |3\rangle\to|4\rangle$, respectively, where $|1\rangle$, $|2\rangle$, $|3\rangle$ can be chosen from $\mathrm{5^2S_{1/2}}$ state of $^{87}\rm{Rb}$, such as $|F=1, m_F =\pm 1\rangle$ and $|F=2, m_F =1\rangle$, while $|4\rangle$ can be selected from $\mathrm{5^2P_{1/2}}$ state, such as $|F'=1, m_F=0\rangle$.
Here we consider both the signal and pump beams are injected into atomic ensemble along the z-axis. The coupling field is consisted of two groups of orthogonalized paired-beams paraxially propagating along the $z$-direction.

Under the rotating-wave approximation, the density-matrix equations for our proposed 4-level tripod-type atomic system can be expressed as \cite{wang2014two}
\begin{equation}\nonumber
\begin{aligned}
	\dot{\rho}_{11}=&-\frac{i}{2}\Omega_s\rho_{14}+\frac{i}{2}\Omega_s^*\rho_{41}+\Gamma_{41}\rho_{44}\\
	\dot{\rho}_{22}=&-\frac{i}{2}\Omega_c\rho_{24}+\frac{i}{2}\Omega_c^*\rho_{42}+\Gamma_{42}\rho_{44} \\
\end{aligned}
\end{equation}
\begin{align}\label{master rho}
	\dot{\rho}_{33}=&-\frac{i}{2}\Omega_p\rho_{34}+\frac{i}{2}\Omega_p^*\rho_{34}\rho_{43}+\Gamma_{43}\rho_{44} \notag\\
	\dot{\rho}_{44}=&\frac{i}{2}(\Omega_s\rho_{14}-\Omega_s^*\rho_{41})+\frac{i}{2}(\Omega_c\rho_{24}-\Omega_c^*\rho_{42})  \notag \\
	& +\frac{i}{2}(\Omega_p\rho_{34}-\Omega_p^*\rho_{43})
	-\Gamma\rho_{44}  \notag\\
	\dot{\rho}_{21}=&i(\Delta_s-\Delta_c)\rho_{21}+\frac{i}{2}\Omega_c^*\rho_{41}-\frac{i}{2}\Omega_s\rho_{24} \notag\\
	\dot{\rho}_{31}=&i(\Delta_s-\Delta_p)\rho_{31}+\frac{i}{2}\Omega_p^*\rho_{41}-\frac{i}{2}\Omega_s\rho_{34}  \notag\\
	\dot{\rho}_{41}=&\frac{i}{2}\Omega_s\rho_{11}+\frac{i}{2}\Omega_c\rho_{21}+\frac{i}{2}\Omega_p\rho_{31}-\frac{i}{2}\Omega_s\rho_{44}  \notag\\
	&+i\Delta_s \rho_{41} - \frac{\Gamma}{2}\rho_{41} \notag \\
	\dot{\rho}_{32}=&i(\Delta_c-\Delta_p)\rho_{32}+\frac{i}{2}\Omega_p^*\rho_{42}-\frac{i}{2}\Omega_c\rho_{34} \notag \\
	\dot{\rho}_{42}=&\frac{i}{2}\Omega_s\rho_{12}+\frac{i}{2}\Omega_c\rho_{22}+\frac{i}{2}\Omega_p\rho_{32}-\frac{i}{2}\Omega_c\rho_{44} \notag \\
	&+i\Delta_c \rho_{42} - \frac{\Gamma}{2}\rho_{42}  \notag \\
	\dot{\rho}_{43}=&\frac{i}{2}\Omega_s\rho_{13}+\frac{i}{2}\Omega_c\rho_{23}+\frac{i}{2}\Omega_p\rho_{33}-\frac{i}{2}\Omega_p\rho_{44} \notag\\
	 &+i\Delta_p \rho_{43}- \frac{\Gamma}{2}\rho_{43}	
\end{align}
where $\Gamma_{nm}$ is the natural decay rate between level $|n\rangle$ and $|m\rangle$ and $\Gamma=\Gamma_{41}+\Gamma_{42}+\Gamma_{43}$. $\Omega_s=\mu_{41}E_s/\hbar$, $\Omega_c=\mu_{42}E_c/\hbar$ and $\Omega_p=\mu_{43}E_p/\hbar$ are Rabi frequencies of signal, coupling and pump fields, where $\mu_{ij}$ is the electric dipole matrix element related to the atomic transition between $|i\rangle$ and $|j\rangle$. $E_{s(c,p)}$ is the strength of corresponding electric field. $\Delta_s=\omega_s-\omega_{14}$, $\Delta_c=\omega_c-\omega_{24}$ and $\Delta_p=\omega_p-\omega_{34}$ denote the frequency detunings. Since the signal field is much weaker than both coupling and pump fields, from Eq~.\eqref{master rho} we can obtain the following relations
\begin{align}\label{rho21 rho31}
		\rho_{21}&=\frac{-\Omega_c^*/2}{\Delta_s-\Delta_c} \rho_{41} \notag \\
		\rho_{31}&=\frac{-\Omega_p^*/2}{\Delta_s-\Delta_p} \rho_{41}.
\end{align}
Substituting Eq~.\eqref{rho21 rho31} into Eq~.\eqref{master rho}, $\rho_{41}$ can be solved as
\begin{equation}\label{rho41}
	\rho_{41}=-\left[ (\Delta_s+\frac{i\Gamma}{2})+\frac{|\Omega_c|^2/4}{\Delta_c-\Delta_s}+\frac{|\Omega_p|^2/4}{\Delta_p-\Delta_s} \right]^{-1} \frac{\Omega_s}{2}
\end{equation}
The susceptibility of atomic medium can be determined through the following formula
\begin{equation}\label{chi}
	\begin{aligned}
		\chi=&2N\mu_{14}\rho_{41}/\epsilon_0E_s \\
		=&\frac{-N|\mu_{14}|^2}{\epsilon_0}\left[ (\Delta_s+\frac{i\Gamma}{2})+\frac{|\Omega_c|^2/4}{\Delta_c-\Delta_s}+\frac{|\Omega_p|^2/4}{\Delta_p-\Delta_s} \right]^{-1}
	\end{aligned}
\end{equation}
where $N$ is the atomic density. The refractive index can be obtained via the relation $n=\sqrt{1+\chi}\approx 1+\chi/2$. From Eq~.\eqref{chi}, one can notice that the spatial profile of the susceptibility is highly dependent on the distribution of the coupling fields, and thus can produce various structures by shaping them. To show that, here we consider that the coupling fields are consisted of two groups of orthogonalized paired-standing waves paraxially propagating along the z-axis (captured by a small angle $\phi$ to the z-axis), as depicted in Fig.\ref{fig1}(b) by the solid and dashed lines, respectively.

\begin{figure}[t]
	\includegraphics[width=0.54\textwidth]{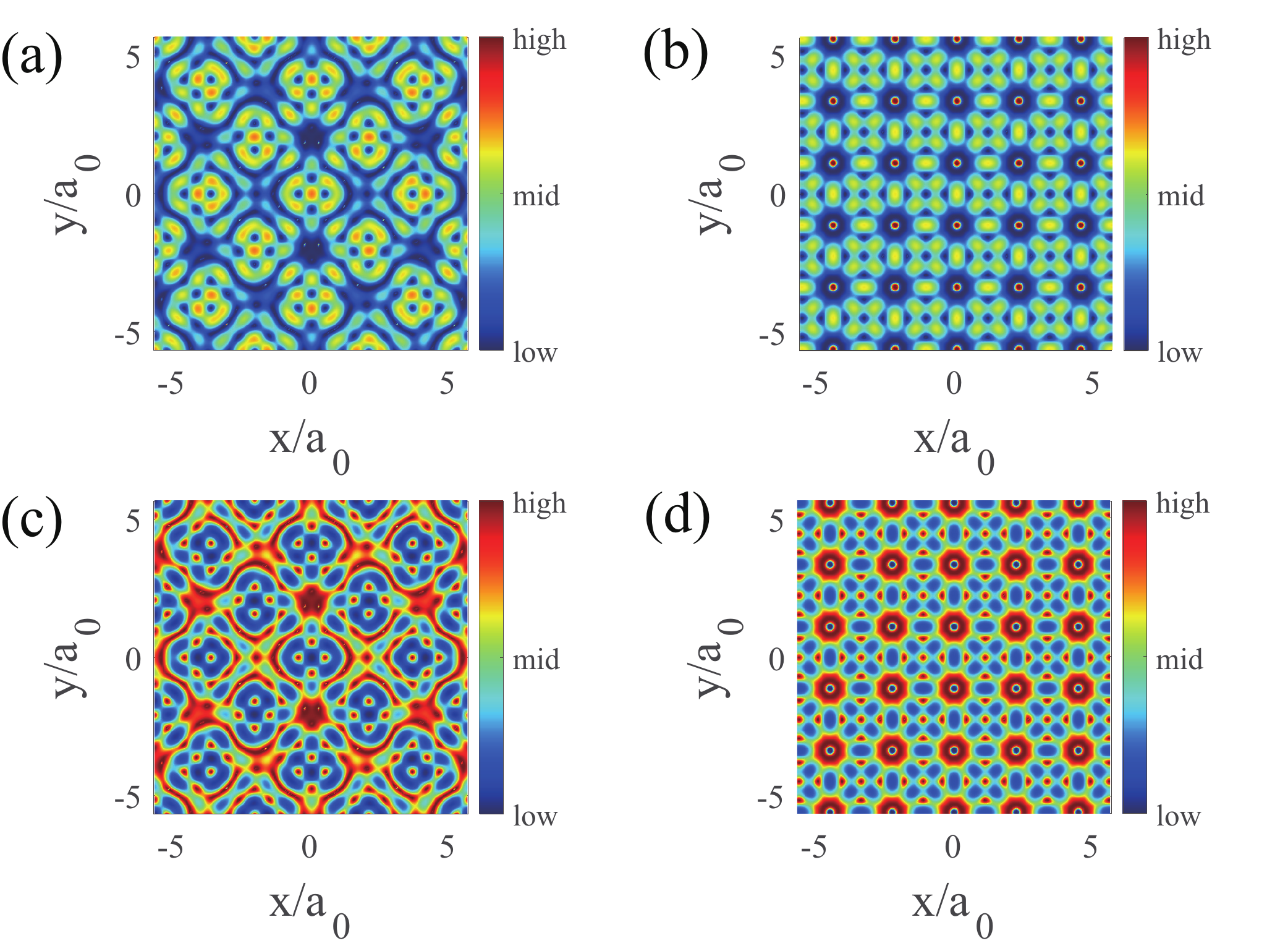}
	\caption{
		Spatial structure of the refractive index lattice. (a) and (c) show the real and imaginary part of the refractive index lattice with $\theta=\pi/6$, which forms an aperiodic moir\'{e} pattern. For comparison, a periodic structure is also shown in (b) and (d) for the real part and imaginary part with $\theta=\arctan 4/3$. Here $a_0=\lambda_c/\sin\phi$ and $\alpha=1$.
	}
	\label{fig2}
\end{figure}
\begin{figure*}[htb]
	\centering
	\includegraphics[width=0.98\textwidth]{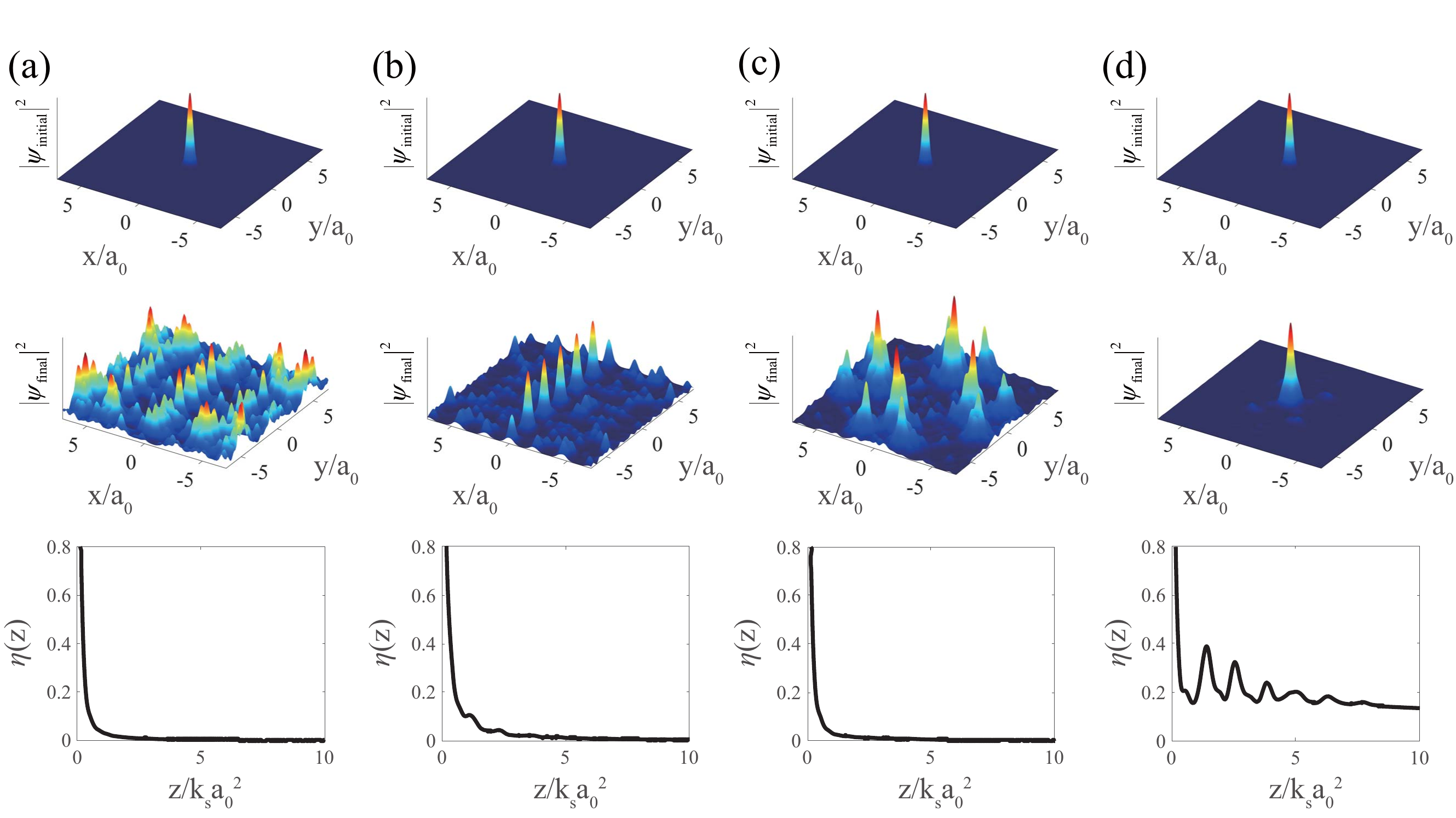}
	\caption{The propagation of light in moir\'{e}  photonic lattice. In (a) and (b), it is shown that when $\theta$ is chosen as a Pythagorean angle $\theta =\arctan 4/3$, the initial Gaussion wavepacket (top row) displays the delocalized behavior for any amplitude ratio $\alpha$ of two superimposed periodic patterns. Here, we choose $\alpha=0.1$ and $\alpha=1$ in (a) and (b), respectively. Such delocalized behavior can also be captured by the vanished IPR with the long propagation distance, as shown in the bottom row of (a) and (b). In (c) and (d), when $\theta=\pi/6$, we find that there is a threshold of $\alpha$. Below that threshold, as shown in the middle panel of (c) ($\alpha=0.1$), the light propagation still shows the delocalized behavior. While above that threshold, as shown in the middle row of (d) ($\alpha=1$), the light propagation will still keep being localized, which can be verified by the non-zero IPR as shown in the bottom row of (d). In the middle row, the propagation distance is chosen as $z/k_sa_0^2=10$. Other parameters are chosen as the same in Fig.2.}
	\label{fig3}
\end{figure*}

The two groups of orthogonalized paired-standing waves can form two superimposed square patterns.  And the total spatial pattern is highly tunable through changing the twisted angle $\theta$ as shown in Fig.\ref{fig1}(b). To be more specific, the standing waves as shown in Fig.\ref{fig1}(b) can be expressed as
\begin{align}
		\vec{E}_{c1}(\vec{r},t)&=E_c'\cos k_0x [ e^{i(k_zz-\omega_c t)}\hat{x}+e^{i(k_zz-\omega_c t-\pi/2)}\hat{y} ] \notag  \\
		\vec{E}_{c2}(\vec{r},t)&=E_c'\cos k_0y [ e^{i(k_zz-\omega_c t+\pi/2)}\hat{x}+e^{i(k_zz-\omega_c t)}\hat{y} ] \notag \\
		\vec{E}_{c3}(\vec{r},t)&=E_c''\cos k_0x' [ e^{i(k_zz-\omega_c t)}\hat{x}'+e^{i(k_zz-\omega_c t-\pi/2)}\hat{y}' ] \notag  \\
		\vec{E}_{c4}(\vec{r},t)&=E_c''\cos k_0y' [ e^{i(k_zz-\omega_c t+\pi/2)}\hat{x}'+e^{i(k_zz-\omega_c t)}\hat{y}' ]
\end{align}
where $k_z=k_c\cos\phi$ and $k_0=k_c\sin\phi$. Unit vectors $\hat{x}',\hat{y}'$ are related to $\hat{x},\hat{y}$ via the relation $\left[x',y'\right]^\mathsf{T}=S\cdot\left[x,y\right]^\mathsf{T}$, where $S=[\cos\theta, -\sin\theta; \sin\theta, \cos\theta]$ is the operator of two dimensional rotation. Therefore, the intensity of coupling field can be expressed as
\begin{align}\label{Ec}
\left|E_c(x,y)\right|^2
=&2E_c'^2 \Big| (\cos k_0y\cdot \hat{x} + \cos k_0x\cdot\hat{y}) \notag \\
+& \alpha(\cos k_0y'\cdot\hat{x}' + \cos k_0 x'\cdot\hat{y}') \Big|^2
\end{align}
where $\alpha=E_c''/E_c'$. As shown in Fig.\ref{fig2}, when varying the twisted angle $\theta$ and amplitude ratio $\alpha$, the interference of coupling fields will produce different spatial pattern and induce an effective 2D photonic lattice in $xy$ plane.  For instance, the periodic structure of refractive index lattice is produced when $\theta$ is a Pythagorean angle, e.g., $\theta=\arctan 4/3$ (see Fig.\ref{fig2} (b) and (d)), otherwise, the aperiodic structure is induced, e.g., $\theta=\pi/6$ (see Fig.\ref{fig2} (a) and (c))

\section{Localization-delocalization Transition}
In the following, we will demonstrate the effect of the spatial profiles of our proposed refractive index lattice through investigating the light propagation within it. The propagation of signal beam $\vec{E}_s (\vec{r},t)$ in the atomic medium is governed by the following electric field wave equation
\begin{equation}
	\nabla^2\vec{E}_s+\frac{\omega_s^2}{c^2}\epsilon(\vec{r})\vec{E}_s=0
\end{equation}
where $\epsilon=1+\chi$ is the relative dielectric constant. We then rewrite $\vec{E}_s (\vec{r},t)$ as $\vec{E}_s (\vec{r},t)=\psi(\vec{r})[\exp{(ik_sz-i\omega_st)}\hat{x}+\exp{(ik_sz-i\omega_st+\pi/2)}\hat{y}]$ with $\psi(\vec{r})$ being the field amplitude. Then, from Eq. (7) a Schr\"{o}dinger-type equation of $\psi(\vec{r})$ can be obtained
\begin{equation}
i\frac{\partial}{\partial z}\psi=-\frac{1}{2k_s}\left( \frac{\partial^2}{\partial x^2}+\frac{\partial^2}{\partial y^2} \right)\psi - \frac{k_s\Delta n(x,y)}{n_0}\psi
\end{equation}
where $\Delta n(x,y)\approx \chi/2$ and $k_s=2\pi n_0/\lambda_s$ being the wave vector of signal beam. $n_0$ is ambient refractive index.

\begin{figure}[htb]
	\centering
	\includegraphics[width=0.4\textwidth]{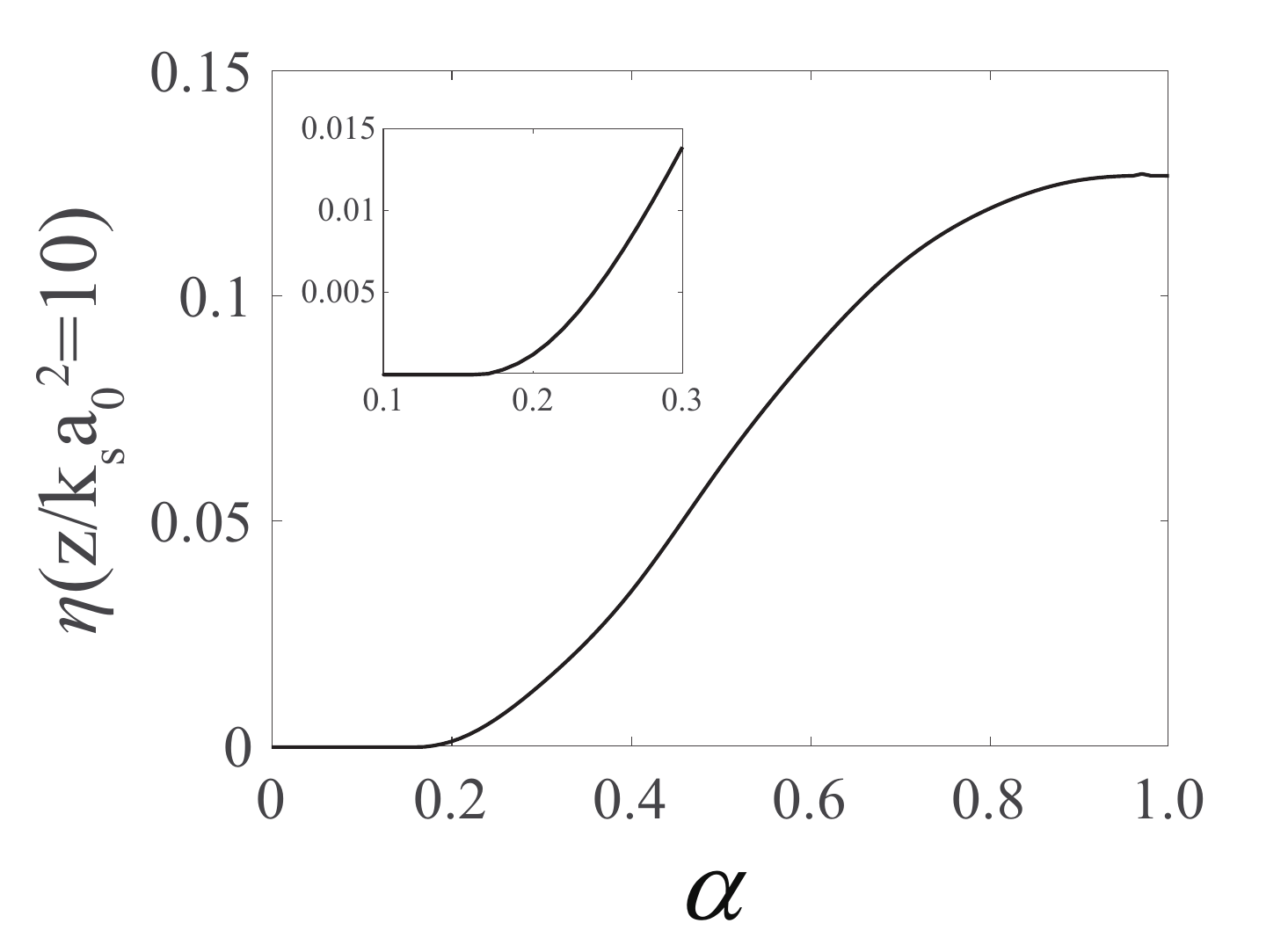}
	\caption{IPR as a function of the amplitude ratio $\alpha$ of two superposed pattern for aperiodic moir\'{e} lattice. A threshold of amplitude ratio $\alpha$  can be determined by the non-zero point of IPR. Other parameters are chosen as the same in Fig.3(d).}
	\label{fig4}
\end{figure}

To investigate the light propagation in the refractive index lattices as shown in Fig.\ref{fig2}, we initialize the signal beam as a Gaussion wavepacket and numerically simulate its propagation.
As shown in Fig.\ref{fig3}, when $\theta$ is chosen as a Pythagorean angle,  for instance, $\theta=\arctan 4/3$, the refractive index lattices possess a spatially periodic structure and the initial Gaussion wavepacket displays the delocalization behavior for arbitrary amplitude ratio $\alpha$ of the two superimposed periodic patterns. When the refractive index lattices possess a spatial
aperiodic structure, for instance, $\theta=\pi/6$, we find that there is a threshold of $\alpha$. Below that threshold, as shown in Fig.\ref{fig3} (c), the light propagation still shows the delocalization behavior. However, if $\alpha$ exceeds the threshold, as shown in Fig.\ref{fig3} (d), the signal beam turns out to be localized. Therefore, there is a localization-delocalization transition (LDT) in aperiodic moir\'{e} lattice, when tuning the amplitudes of the two superimposed periodic patterns.
To quantitatively analyze the LDT here, we introduce the factor inverse participation ratio (IPR)\cite{evers2000fluctuations} defined as $\eta(z)=\int |\psi(\vec{r})|^4 dxdy /\left( \int |\psi(\vec{r})|^2 dxdy \right)^{2} $. The localized behavior can be captured by the non-zero IPR. As shown in Fig.\ref{fig4}, the threshold of amplitude ratio in the aperiodic moir\'{e} lattice separating two distinct regimes in the LDT can be determined by
the non-zero point of IPR when varying $\alpha$.

\begin{figure}[b]
	\centering
	\includegraphics[width=0.5\textwidth]{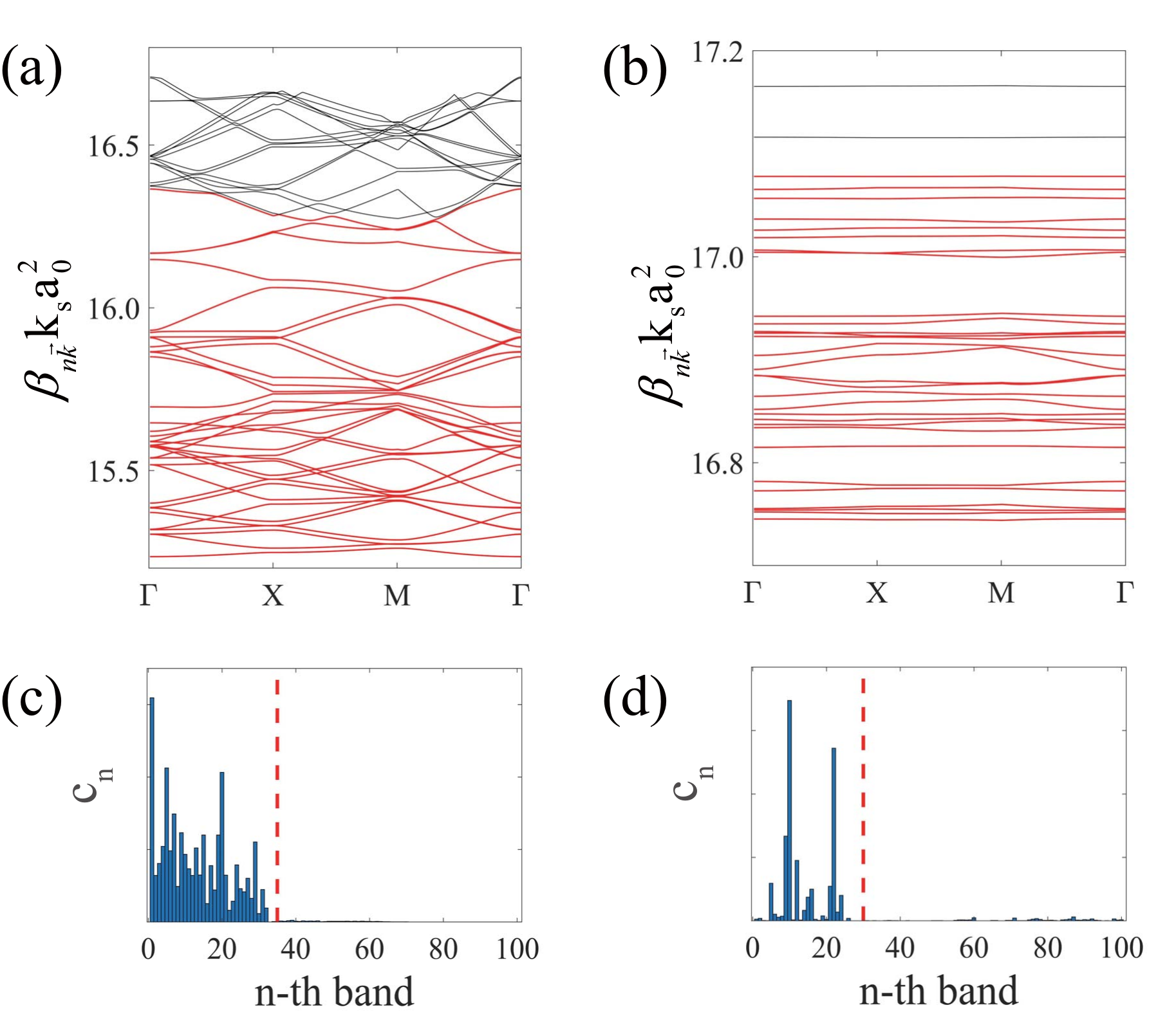}
	\caption{
		(a) and (b) The single-particle dispersion relation of aperiodic lattice with $\theta=\pi/6$ under the Pythagorean approximation via choosing $\theta= \arctan 120/209$ for $\alpha=0.1$ and $\alpha=1.0$, respectively. (c) and (d) The band occupation probability $c_n$ defined in the main text for the case with $\alpha=0.1$ and $\alpha=1.0$, respectively. In (a) and (b), the red colored bands are the occupied bands of the chosen initial wavepacket.
	}
	\label{fig5}
\end{figure}

To understand the localization of light in aperiodic moir\'{e} lattice, we calculate its single-particle dispersion relation through
approximating the non-Pythagorean twist angle by a Pythagorean one \cite{wang2020localization}. For instance, here we use $\theta=\arctan{(120/209)}$ to approximate $\theta=\pi/6$. Under such an approximation, the single-particle dispersion of aperiodic lattice can be obtained by the plane-wave expansion method through introducing the Bloch basis $\psi_{n\vec{k}}=\sum_{\vec{G}} u_{n\vec{k},\vec{G}}|\vec{k}+\vec{G}\rangle$ with the Bloch vector $\vec{k}$ and reciprocal lattice vector $\vec{G}$. Here $n$ labels the band index.
Then, the single-particle dispersion relation can be obtained through solving the eigen-problem via the following relation
\begin{equation}
	\begin{aligned}
		\frac{(\vec{k}+\vec{G})^2}{2k_s}u_{n\vec{k},\vec{G}} -\frac{k_s}{n_0} \sum_{\vec{G'}}\langle \vec{k}+\vec{G} | \Delta n(x,y) | \vec{k}+\vec{G'}\rangle u_{n\vec{k},\vec{G'}}& \\
		= \beta_{n\vec{k}} u_{n\vec{k},\vec{G}}&
	\end{aligned}	
\end{equation}
where $\beta_{n\vec{k}}$ is the dispersion relation of 2D Bloch waves. As shown in Fig.\ref{fig5}, when the amplitude ratio $\alpha$ increases, more lower bands become extremely flat. Since the flat bands support quasi-nondiffracting localized modes, the initially localized beam launched into such moir\'{e} lattice will remain localized. To show that, we define a quantity $c_n\equiv\langle \psi_n | \psi_0\rangle$ with $|\psi_0\rangle$ labeling the initial Gaussion wavepacket and $|\psi_n\rangle$ standing for the eigenstates solved from Eq. (9), to decompose the initial state into the eigenstates of aperiodic lattice. It can capture the band occupation probability of the chosen initial state.
For instance, as shown in Fig.\ref{fig5} (a) and (c), when the amplitude ratio $\alpha$ is below the threshold, the occupied bands of the initial wavepacket ($c_n\ne 0$) are dispersive. Therefore, the light propagator presents the delocalized behavior. While increasing $\alpha$ above the threshold, the occupied bands of the initial wavepacket are flat. Therefore, such flat bands drive the LDT in aperiodic moir\'{e} lattice, since the flat bands support quasi-nondiffracting localized modes.
%\bigskip
\section{Conclusion}
In summary, we propose a four-level tripod-type EIT scheme in atomic ensembles to induce a photonic moir\'{e} lattice. Such a lattice shows great tunability of changing the spatial structure. Both periodic and  aperiodic structures can be achieved. We further explore the LDT behavior in the aperiodic moir\'{e} lattice through investigating the light propagation. A threshold of amplitude ratio of two superposed patterns has been found. Such a phenomenon can be understood through analysis of the flat-band physics of moir\'{e} lattice. Our proposal would provide a promising approach to manipulate the light propagation through electromagnetically induced photonic lattices and thus have potential applications in optical information techniques.

\section{Acknowledgement}
This work is supported by the National Key R$\&$D Program of China (2021YFA1401700), NSFC (Grants No. 12074305, 12147137, 11774282), the National Key Research and Development Program of China (2018YFA0307600), Xiaomi Young Scholar Program (R. T., S. L., M. A. and B. L.), and Shaanxi Academy of Fundamental Sciences (Mathematics, Physics) (Grant No. 22JSY036), Xi'an Jiaotong University through the Young Top Talents Support Plan, Basic Research Funding (Grant No. xtr042021012) (Z. C.). We also thank the HPC platform of Xi'An Jiaotong University, where our numerical calculations was performed.

\bibliographystyle{apsrev}
\bibliography{moire-ref}

\begin{thebibliography}{27}
\expandafter\ifx\csname natexlab\endcsname\relax\def\natexlab#1{#1}\fi
\expandafter\ifx\csname bibnamefont\endcsname\relax
  \def\bibnamefont#1{#1}\fi
\expandafter\ifx\csname bibfnamefont\endcsname\relax
  \def\bibfnamefont#1{#1}\fi
\expandafter\ifx\csname citenamefont\endcsname\relax
  \def\citenamefont#1{#1}\fi
\expandafter\ifx\csname url\endcsname\relax
  \def\url#1{\texttt{#1}}\fi
\expandafter\ifx\csname urlprefix\endcsname\relax\def\urlprefix{URL }\fi
\providecommand{\bibinfo}[2]{#2}
\providecommand{\eprint}[2][]{\url{#2}}

\bibitem[{\citenamefont{Akahane et~al.}(2003)\citenamefont{Akahane, Asano,
  Song, and Noda}}]{akahane2003high}
\bibinfo{author}{\bibfnamefont{Y.}~\bibnamefont{Akahane}},
  \bibinfo{author}{\bibfnamefont{T.}~\bibnamefont{Asano}},
  \bibinfo{author}{\bibfnamefont{B.-S.} \bibnamefont{Song}}, \bibnamefont{and}
  \bibinfo{author}{\bibfnamefont{S.}~\bibnamefont{Noda}},
  \bibinfo{journal}{Nature} \textbf{\bibinfo{volume}{425}},
  \bibinfo{pages}{944} (\bibinfo{year}{2003}).

\bibitem[{\citenamefont{Park et~al.}(2004)\citenamefont{Park, Kim, Kwon, Ju,
  Yang, Baek, Kim, and Lee}}]{park2004electrically}
\bibinfo{author}{\bibfnamefont{H.-G.} \bibnamefont{Park}},
  \bibinfo{author}{\bibfnamefont{S.-H.} \bibnamefont{Kim}},
  \bibinfo{author}{\bibfnamefont{S.-H.} \bibnamefont{Kwon}},
  \bibinfo{author}{\bibfnamefont{Y.-G.} \bibnamefont{Ju}},
  \bibinfo{author}{\bibfnamefont{J.-K.} \bibnamefont{Yang}},
  \bibinfo{author}{\bibfnamefont{J.-H.} \bibnamefont{Baek}},
  \bibinfo{author}{\bibfnamefont{S.-B.} \bibnamefont{Kim}}, \bibnamefont{and}
  \bibinfo{author}{\bibfnamefont{Y.-H.} \bibnamefont{Lee}},
  \bibinfo{journal}{Science} \textbf{\bibinfo{volume}{305}},
  \bibinfo{pages}{1444} (\bibinfo{year}{2004}).

\bibitem[{\citenamefont{Joannopoulos et~al.}(2008)\citenamefont{Joannopoulos,
  Johnson, Winn, and Meade}}]{joannopoulos2008molding}
\bibinfo{author}{\bibfnamefont{J.~D.} \bibnamefont{Joannopoulos}},
  \bibinfo{author}{\bibfnamefont{S.~G.} \bibnamefont{Johnson}},
  \bibinfo{author}{\bibfnamefont{J.~N.} \bibnamefont{Winn}}, \bibnamefont{and}
  \bibinfo{author}{\bibfnamefont{R.~D.} \bibnamefont{Meade}},
  \bibinfo{journal}{Princeton Univ. Press, Princeton, NJ}
  (\bibinfo{year}{2008}).

\bibitem[{\citenamefont{Smith et~al.}(2004)\citenamefont{Smith, Pendry, and
  Wiltshire}}]{smith2004metamaterials}
\bibinfo{author}{\bibfnamefont{D.~R.} \bibnamefont{Smith}},
  \bibinfo{author}{\bibfnamefont{J.~B.} \bibnamefont{Pendry}},
  \bibnamefont{and} \bibinfo{author}{\bibfnamefont{M.~C.}
  \bibnamefont{Wiltshire}}, \bibinfo{journal}{Science}
  \textbf{\bibinfo{volume}{305}}, \bibinfo{pages}{788} (\bibinfo{year}{2004}).

\bibitem[{\citenamefont{Schurig et~al.}(2006)\citenamefont{Schurig, Mock,
  Justice, Cummer, Pendry, Starr, and Smith}}]{schurig2006metamaterial}
\bibinfo{author}{\bibfnamefont{D.}~\bibnamefont{Schurig}},
  \bibinfo{author}{\bibfnamefont{J.~J.} \bibnamefont{Mock}},
  \bibinfo{author}{\bibfnamefont{B.}~\bibnamefont{Justice}},
  \bibinfo{author}{\bibfnamefont{S.~A.} \bibnamefont{Cummer}},
  \bibinfo{author}{\bibfnamefont{J.~B.} \bibnamefont{Pendry}},
  \bibinfo{author}{\bibfnamefont{A.~F.} \bibnamefont{Starr}}, \bibnamefont{and}
  \bibinfo{author}{\bibfnamefont{D.~R.} \bibnamefont{Smith}},
  \bibinfo{journal}{Science} \textbf{\bibinfo{volume}{314}},
  \bibinfo{pages}{977} (\bibinfo{year}{2006}).

\bibitem[{\citenamefont{Han et~al.}(2014)\citenamefont{Han, Bai, Thong, Li, and
  Qiu}}]{han2014full}
\bibinfo{author}{\bibfnamefont{T.}~\bibnamefont{Han}},
  \bibinfo{author}{\bibfnamefont{X.}~\bibnamefont{Bai}},
  \bibinfo{author}{\bibfnamefont{J.~T.} \bibnamefont{Thong}},
  \bibinfo{author}{\bibfnamefont{B.}~\bibnamefont{Li}}, \bibnamefont{and}
  \bibinfo{author}{\bibfnamefont{C.-W.} \bibnamefont{Qiu}},
  \bibinfo{journal}{Adv. Mater.} \textbf{\bibinfo{volume}{26}},
  \bibinfo{pages}{1731} (\bibinfo{year}{2014}).

\bibitem[{\citenamefont{Hu et~al.}(2005)\citenamefont{Hu, Liu, Tian, Cheng, and
  Zhang}}]{hu2005ultrafast}
\bibinfo{author}{\bibfnamefont{X.}~\bibnamefont{Hu}},
  \bibinfo{author}{\bibfnamefont{Y.}~\bibnamefont{Liu}},
  \bibinfo{author}{\bibfnamefont{J.}~\bibnamefont{Tian}},
  \bibinfo{author}{\bibfnamefont{B.}~\bibnamefont{Cheng}}, \bibnamefont{and}
  \bibinfo{author}{\bibfnamefont{D.}~\bibnamefont{Zhang}},
  \bibinfo{journal}{Appl. Phys. Lett.} \textbf{\bibinfo{volume}{86}},
  \bibinfo{pages}{121102} (\bibinfo{year}{2005}).

\bibitem[{\citenamefont{Zhang and Liu}(2008)}]{zhang2008superlenses}
\bibinfo{author}{\bibfnamefont{X.}~\bibnamefont{Zhang}} \bibnamefont{and}
  \bibinfo{author}{\bibfnamefont{Z.}~\bibnamefont{Liu}}, \bibinfo{journal}{Nat.
  Mater.} \textbf{\bibinfo{volume}{7}}, \bibinfo{pages}{435}
  (\bibinfo{year}{2008}).

\bibitem[{\citenamefont{Bhandari}(1997)}]{bhandari1997polarization}
\bibinfo{author}{\bibfnamefont{R.}~\bibnamefont{Bhandari}},
  \bibinfo{journal}{Phys. Rep.} \textbf{\bibinfo{volume}{281}},
  \bibinfo{pages}{1} (\bibinfo{year}{1997}).

\bibitem[{\citenamefont{Lu et~al.}(2014)\citenamefont{Lu, Joannopoulos, and
  Solja{\v{c}}i{\'c}}}]{lu2014topological}
\bibinfo{author}{\bibfnamefont{L.}~\bibnamefont{Lu}},
  \bibinfo{author}{\bibfnamefont{J.~D.} \bibnamefont{Joannopoulos}},
  \bibnamefont{and}
  \bibinfo{author}{\bibfnamefont{M.}~\bibnamefont{Solja{\v{c}}i{\'c}}},
  \bibinfo{journal}{Nat. Photonics} \textbf{\bibinfo{volume}{8}},
  \bibinfo{pages}{821} (\bibinfo{year}{2014}).

\bibitem[{\citenamefont{Ozawa et~al.}(2019)\citenamefont{Ozawa, Price, Amo,
  Goldman, Hafezi, Lu, Rechtsman, Schuster, Simon, Zilberberg
  et~al.}}]{ozawa2019topological}
\bibinfo{author}{\bibfnamefont{T.}~\bibnamefont{Ozawa}},
  \bibinfo{author}{\bibfnamefont{H.~M.} \bibnamefont{Price}},
  \bibinfo{author}{\bibfnamefont{A.}~\bibnamefont{Amo}},
  \bibinfo{author}{\bibfnamefont{N.}~\bibnamefont{Goldman}},
  \bibinfo{author}{\bibfnamefont{M.}~\bibnamefont{Hafezi}},
  \bibinfo{author}{\bibfnamefont{L.}~\bibnamefont{Lu}},
  \bibinfo{author}{\bibfnamefont{M.~C.} \bibnamefont{Rechtsman}},
  \bibinfo{author}{\bibfnamefont{D.}~\bibnamefont{Schuster}},
  \bibinfo{author}{\bibfnamefont{J.}~\bibnamefont{Simon}},
  \bibinfo{author}{\bibfnamefont{O.}~\bibnamefont{Zilberberg}},
  \bibnamefont{et~al.}, \bibinfo{journal}{Rev. Mod. Phys.}
  \textbf{\bibinfo{volume}{91}}, \bibinfo{pages}{015006}
  (\bibinfo{year}{2019}).

\bibitem[{\citenamefont{Ling et~al.}(1998)\citenamefont{Ling, Li, and
  Xiao}}]{ling1998electromagnetically}
\bibinfo{author}{\bibfnamefont{H.~Y.} \bibnamefont{Ling}},
  \bibinfo{author}{\bibfnamefont{Y.-Q.} \bibnamefont{Li}}, \bibnamefont{and}
  \bibinfo{author}{\bibfnamefont{M.}~\bibnamefont{Xiao}},
  \bibinfo{journal}{Phys. Lett. A} \textbf{\bibinfo{volume}{57}},
  \bibinfo{pages}{1338} (\bibinfo{year}{1998}).

\bibitem[{\citenamefont{Sheng et~al.}(2015)\citenamefont{Sheng, Wang, Miri,
  Christodoulides, and Xiao}}]{sheng2015observation}
\bibinfo{author}{\bibfnamefont{J.}~\bibnamefont{Sheng}},
  \bibinfo{author}{\bibfnamefont{J.}~\bibnamefont{Wang}},
  \bibinfo{author}{\bibfnamefont{M.-A.} \bibnamefont{Miri}},
  \bibinfo{author}{\bibfnamefont{D.~N.} \bibnamefont{Christodoulides}},
  \bibnamefont{and} \bibinfo{author}{\bibfnamefont{M.}~\bibnamefont{Xiao}},
  \bibinfo{journal}{Opt. Express} \textbf{\bibinfo{volume}{23}},
  \bibinfo{pages}{19777} (\bibinfo{year}{2015}).

\bibitem[{\citenamefont{Zhang et~al.}(2018)\citenamefont{Zhang, Feng, Liu,
  Sheng, Zhang, Zhang, and Xiao}}]{zhang2018controllable}
\bibinfo{author}{\bibfnamefont{Z.}~\bibnamefont{Zhang}},
  \bibinfo{author}{\bibfnamefont{J.}~\bibnamefont{Feng}},
  \bibinfo{author}{\bibfnamefont{X.}~\bibnamefont{Liu}},
  \bibinfo{author}{\bibfnamefont{J.}~\bibnamefont{Sheng}},
  \bibinfo{author}{\bibfnamefont{Y.}~\bibnamefont{Zhang}},
  \bibinfo{author}{\bibfnamefont{Y.}~\bibnamefont{Zhang}}, \bibnamefont{and}
  \bibinfo{author}{\bibfnamefont{M.}~\bibnamefont{Xiao}},
  \bibinfo{journal}{Opt. Lett.} \textbf{\bibinfo{volume}{43}},
  \bibinfo{pages}{919} (\bibinfo{year}{2018}).

\bibitem[{\citenamefont{Yuan et~al.}(2019)\citenamefont{Yuan, Wu, Li, Wang,
  Zhang, Xiao, and Jia}}]{yuan2019integer}
\bibinfo{author}{\bibfnamefont{J.}~\bibnamefont{Yuan}},
  \bibinfo{author}{\bibfnamefont{C.}~\bibnamefont{Wu}},
  \bibinfo{author}{\bibfnamefont{Y.}~\bibnamefont{Li}},
  \bibinfo{author}{\bibfnamefont{L.}~\bibnamefont{Wang}},
  \bibinfo{author}{\bibfnamefont{Y.}~\bibnamefont{Zhang}},
  \bibinfo{author}{\bibfnamefont{L.}~\bibnamefont{Xiao}}, \bibnamefont{and}
  \bibinfo{author}{\bibfnamefont{S.}~\bibnamefont{Jia}}, \bibinfo{journal}{Opt.
  Express} \textbf{\bibinfo{volume}{27}}, \bibinfo{pages}{92}
  (\bibinfo{year}{2019}).

\bibitem[{\citenamefont{Radwell et~al.}(2015)\citenamefont{Radwell, Clark,
  Piccirillo, Barnett, and Franke-Arnold}}]{radwell2015spatially}
\bibinfo{author}{\bibfnamefont{N.}~\bibnamefont{Radwell}},
  \bibinfo{author}{\bibfnamefont{T.~W.} \bibnamefont{Clark}},
  \bibinfo{author}{\bibfnamefont{B.}~\bibnamefont{Piccirillo}},
  \bibinfo{author}{\bibfnamefont{S.~M.} \bibnamefont{Barnett}},
  \bibnamefont{and}
  \bibinfo{author}{\bibfnamefont{S.}~\bibnamefont{Franke-Arnold}},
  \bibinfo{journal}{Phys. Rev. Lett.} \textbf{\bibinfo{volume}{114}},
  \bibinfo{pages}{123603} (\bibinfo{year}{2015}).

\bibitem[{\citenamefont{Yang et~al.}(2020)\citenamefont{Yang, Zhang, Zhang, and
  Wu}}]{yang2020dynamically}
\bibinfo{author}{\bibfnamefont{H.}~\bibnamefont{Yang}},
  \bibinfo{author}{\bibfnamefont{T.}~\bibnamefont{Zhang}},
  \bibinfo{author}{\bibfnamefont{Y.}~\bibnamefont{Zhang}}, \bibnamefont{and}
  \bibinfo{author}{\bibfnamefont{J.-H.} \bibnamefont{Wu}},
  \bibinfo{journal}{Phys. Lett. A} \textbf{\bibinfo{volume}{101}},
  \bibinfo{pages}{053856} (\bibinfo{year}{2020}).

\bibitem[{\citenamefont{Fleischer et~al.}(2003)\citenamefont{Fleischer, Carmon,
  Segev, Efremidis, and Christodoulides}}]{fleischer2003observation}
\bibinfo{author}{\bibfnamefont{J.~W.} \bibnamefont{Fleischer}},
  \bibinfo{author}{\bibfnamefont{T.}~\bibnamefont{Carmon}},
  \bibinfo{author}{\bibfnamefont{M.}~\bibnamefont{Segev}},
  \bibinfo{author}{\bibfnamefont{N.~K.} \bibnamefont{Efremidis}},
  \bibnamefont{and} \bibinfo{author}{\bibfnamefont{D.~N.}
  \bibnamefont{Christodoulides}}, \bibinfo{journal}{Phys. Rev. Lett.}
  \textbf{\bibinfo{volume}{90}}, \bibinfo{pages}{023902}
  (\bibinfo{year}{2003}).

\bibitem[{\citenamefont{Michinel et~al.}(2006)\citenamefont{Michinel,
  Paz-Alonso, and P{\'e}rez-Garc{\'\i}a}}]{michinel2006turning}
\bibinfo{author}{\bibfnamefont{H.}~\bibnamefont{Michinel}},
  \bibinfo{author}{\bibfnamefont{M.~J.} \bibnamefont{Paz-Alonso}},
  \bibnamefont{and} \bibinfo{author}{\bibfnamefont{V.~M.}
  \bibnamefont{P{\'e}rez-Garc{\'\i}a}}, \bibinfo{journal}{Phys. Rev. Lett.}
  \textbf{\bibinfo{volume}{96}}, \bibinfo{pages}{023903}
  (\bibinfo{year}{2006}).

\bibitem[{\citenamefont{Zhang et~al.}(2011)\citenamefont{Zhang, Wang, Nie, Li,
  Chen, Lu, and Xiao}}]{zhang2011four}
\bibinfo{author}{\bibfnamefont{Y.}~\bibnamefont{Zhang}},
  \bibinfo{author}{\bibfnamefont{Z.}~\bibnamefont{Wang}},
  \bibinfo{author}{\bibfnamefont{Z.}~\bibnamefont{Nie}},
  \bibinfo{author}{\bibfnamefont{C.}~\bibnamefont{Li}},
  \bibinfo{author}{\bibfnamefont{H.}~\bibnamefont{Chen}},
  \bibinfo{author}{\bibfnamefont{K.}~\bibnamefont{Lu}}, \bibnamefont{and}
  \bibinfo{author}{\bibfnamefont{M.}~\bibnamefont{Xiao}},
  \bibinfo{journal}{Phys. Rev. Lett.} \textbf{\bibinfo{volume}{106}},
  \bibinfo{pages}{093904} (\bibinfo{year}{2011}).

\bibitem[{\citenamefont{Longo et~al.}(2010)\citenamefont{Longo, Schmitteckert,
  and Busch}}]{longo2010few}
\bibinfo{author}{\bibfnamefont{P.}~\bibnamefont{Longo}},
  \bibinfo{author}{\bibfnamefont{P.}~\bibnamefont{Schmitteckert}},
  \bibnamefont{and} \bibinfo{author}{\bibfnamefont{K.}~\bibnamefont{Busch}},
  \bibinfo{journal}{Phys. Rev. Lett.} \textbf{\bibinfo{volume}{104}},
  \bibinfo{pages}{023602} (\bibinfo{year}{2010}).

\bibitem[{\citenamefont{Rechtsman et~al.}(2013)\citenamefont{Rechtsman, Zeuner,
  Plotnik, Lumer, Podolsky, Dreisow, Nolte, Segev, and
  Szameit}}]{rechtsman2013photonic}
\bibinfo{author}{\bibfnamefont{M.~C.} \bibnamefont{Rechtsman}},
  \bibinfo{author}{\bibfnamefont{J.~M.} \bibnamefont{Zeuner}},
  \bibinfo{author}{\bibfnamefont{Y.}~\bibnamefont{Plotnik}},
  \bibinfo{author}{\bibfnamefont{Y.}~\bibnamefont{Lumer}},
  \bibinfo{author}{\bibfnamefont{D.}~\bibnamefont{Podolsky}},
  \bibinfo{author}{\bibfnamefont{F.}~\bibnamefont{Dreisow}},
  \bibinfo{author}{\bibfnamefont{S.}~\bibnamefont{Nolte}},
  \bibinfo{author}{\bibfnamefont{M.}~\bibnamefont{Segev}}, \bibnamefont{and}
  \bibinfo{author}{\bibfnamefont{A.}~\bibnamefont{Szameit}},
  \bibinfo{journal}{Nature} \textbf{\bibinfo{volume}{496}},
  \bibinfo{pages}{196} (\bibinfo{year}{2013}).

\bibitem[{\citenamefont{Peruzzo et~al.}(2010)\citenamefont{Peruzzo, Lobino,
  Matthews, Matsuda, Politi, Poulios, Zhou, Lahini, Ismail, W{\"o}rhoff
  et~al.}}]{peruzzo2010quantum}
\bibinfo{author}{\bibfnamefont{A.}~\bibnamefont{Peruzzo}},
  \bibinfo{author}{\bibfnamefont{M.}~\bibnamefont{Lobino}},
  \bibinfo{author}{\bibfnamefont{J.~C.} \bibnamefont{Matthews}},
  \bibinfo{author}{\bibfnamefont{N.}~\bibnamefont{Matsuda}},
  \bibinfo{author}{\bibfnamefont{A.}~\bibnamefont{Politi}},
  \bibinfo{author}{\bibfnamefont{K.}~\bibnamefont{Poulios}},
  \bibinfo{author}{\bibfnamefont{X.-Q.} \bibnamefont{Zhou}},
  \bibinfo{author}{\bibfnamefont{Y.}~\bibnamefont{Lahini}},
  \bibinfo{author}{\bibfnamefont{N.}~\bibnamefont{Ismail}},
  \bibinfo{author}{\bibfnamefont{K.}~\bibnamefont{W{\"o}rhoff}},
  \bibnamefont{et~al.}, \bibinfo{journal}{Science}
  \textbf{\bibinfo{volume}{329}}, \bibinfo{pages}{1500} (\bibinfo{year}{2010}).

\bibitem[{\citenamefont{Wang et~al.}(2020)\citenamefont{Wang, Zheng, Chen,
  Huang, Kartashov, Torner, Konotop, and Ye}}]{wang2020localization}
\bibinfo{author}{\bibfnamefont{P.}~\bibnamefont{Wang}},
  \bibinfo{author}{\bibfnamefont{Y.}~\bibnamefont{Zheng}},
  \bibinfo{author}{\bibfnamefont{X.}~\bibnamefont{Chen}},
  \bibinfo{author}{\bibfnamefont{C.}~\bibnamefont{Huang}},
  \bibinfo{author}{\bibfnamefont{Y.~V.} \bibnamefont{Kartashov}},
  \bibinfo{author}{\bibfnamefont{L.}~\bibnamefont{Torner}},
  \bibinfo{author}{\bibfnamefont{V.~V.} \bibnamefont{Konotop}},
  \bibnamefont{and} \bibinfo{author}{\bibfnamefont{F.}~\bibnamefont{Ye}},
  \bibinfo{journal}{Nature} \textbf{\bibinfo{volume}{577}}, \bibinfo{pages}{42}
  (\bibinfo{year}{2020}).

\bibitem[{\citenamefont{Fu et~al.}(2020)\citenamefont{Fu, Wang, Huang,
  Kartashov, Torner, Konotop, and Ye}}]{fu2020optical}
\bibinfo{author}{\bibfnamefont{Q.}~\bibnamefont{Fu}},
  \bibinfo{author}{\bibfnamefont{P.}~\bibnamefont{Wang}},
  \bibinfo{author}{\bibfnamefont{C.}~\bibnamefont{Huang}},
  \bibinfo{author}{\bibfnamefont{Y.~V.} \bibnamefont{Kartashov}},
  \bibinfo{author}{\bibfnamefont{L.}~\bibnamefont{Torner}},
  \bibinfo{author}{\bibfnamefont{V.~V.} \bibnamefont{Konotop}},
  \bibnamefont{and} \bibinfo{author}{\bibfnamefont{F.}~\bibnamefont{Ye}},
  \bibinfo{journal}{Nat. Photonics} \textbf{\bibinfo{volume}{14}},
  \bibinfo{pages}{663} (\bibinfo{year}{2020}).

\bibitem[{\citenamefont{Wang et~al.}(2014)\citenamefont{Wang, Zhou, Hu, Niu,
  and Gong}}]{wang2014two}
\bibinfo{author}{\bibfnamefont{L.}~\bibnamefont{Wang}},
  \bibinfo{author}{\bibfnamefont{F.}~\bibnamefont{Zhou}},
  \bibinfo{author}{\bibfnamefont{P.}~\bibnamefont{Hu}},
  \bibinfo{author}{\bibfnamefont{Y.}~\bibnamefont{Niu}}, \bibnamefont{and}
  \bibinfo{author}{\bibfnamefont{S.}~\bibnamefont{Gong}}, \bibinfo{journal}{J.
  Phys. B: At., Mol. Opt. Phys.} \textbf{\bibinfo{volume}{47}},
  \bibinfo{pages}{225501} (\bibinfo{year}{2014}).

\bibitem[{\citenamefont{Evers and Mirlin}(2000)}]{evers2000fluctuations}
\bibinfo{author}{\bibfnamefont{F.}~\bibnamefont{Evers}} \bibnamefont{and}
  \bibinfo{author}{\bibfnamefont{A.}~\bibnamefont{Mirlin}},
  \bibinfo{journal}{Phys. Rev. Lett.} \textbf{\bibinfo{volume}{84}},
  \bibinfo{pages}{3690} (\bibinfo{year}{2000}).

\end{thebibliography}

\end{document}